\documentclass{article}
\usepackage[T1]{fontenc} % add special characters (e.g., umlaute)
\usepackage[utf8]{inputenc} % set utf-8 as default input encoding
\usepackage{ismir,amsmath,cite,url}
\usepackage{graphicx}
\usepackage{color}
\usepackage[inline]{enumitem}
\usepackage{float}

\usepackage{lineno}
\title{The Music Meta Ontology: a flexible semantic model \\ for the interoperability of music metadata}

\multauthor
{Jacopo de Berardinis$^1$ \hspace{.1cm} Valentina Anita Carriero $^2$ \hspace{.1cm} Albert Meroño-Penuela$^1$}
{ \bfseries{Andrea Poltronieri$^2$ \hspace{.1cm} Valentina Presutti$^2$} \\
$^1$ King's College London, UK\\
$^2$ University of Bologna, Italy\\
{\tt\small jacopo.deberardinis@kcl.ac.uk, andrea.poltronieri2@unibo.it}
}

\def\authorname{J. de Berardinis, V.A. Carriero, A. Meroño-Peñuela, A. Poltronieri, and V. Presutti}
\begin{document}

\maketitle
%
% Music metadata is everywhere, from radio broadcast and streaming services to music databases, libraries, digital catalogues, .
\begin{abstract} % 150-200 words
The semantic description of music metadata is a key requirement for the creation of music datasets that can be aligned, integrated, and accessed for information retrieval and knowledge discovery.
It is nonetheless an open challenge due to the complexity of musical concepts arising from different genres, styles, and periods -- standing to benefit from a lingua franca to accommodate various stakeholders (musicologists, librarians, data engineers, etc.).
To initiate this transition, we introduce the Music Meta ontology, a rich and flexible semantic model to describe music metadata related to artists, compositions, performances, recordings, and links. 
We follow eXtreme Design methodologies and best practices for data engineering, to reflect the perspectives and the requirements of various stakeholders into the design of the model, while leveraging ontology design patterns and accounting for provenance at different levels (claims, links).
After presenting the main features of Music Meta, we provide a first evaluation of the model, alignments to other schema (Music Ontology, DOREMUS, Wikidata), and support for data transformation.
% The reuse of Music Meta promotes interoperability and integration of music data across different domains and applications, making it easier to discover, access, and analyse music information across different sources.
% Our model ensures interoperability and integration of music data across different domains and applications
\end{abstract}

\section{Introduction}\label{sec:intro}

A music analyst, a computational musicologist, a music librarian, and a data engineer are working on a joint project.
They need to contribute data from various musical sources, ranging from music libraries, annotated corpora and tune books, to audiovisual archives, radio broadcasts, and music catalogues.
All data is eventually merged/aggregated as interconnected corpora, and linked to online music databases (e.g. MusicBrainz, Discogs) and knowledge bases (e.g. Wikidata).
This creates opportunities to link cultural heritage artefacts to music industry data (streaming services, music professionals, etc.) and viceversa.
% information retrieval, knowledge discovery, and creative workflows

This plot subsumes a recurring challenge for musical heritage projects \cite{polifonia2021report}.
Besides the individual requirements of each stakeholder -- possibly rooted in different music genres, periods and datasets, a fundamental requirement is the interoperability of music metadata.

Music metadata (alias bibliographic, or documentary music data) is used to consistently identify and describe musical works, their artists, recordings, and performances.
For music industry, it allows for efficient management and distribution of music, which facilitate search and recommendation \cite{pachet2005knowledge}.
When metadata is accurate, it ensures that artists receive proper credit and compensation \cite{sitonio2018impact}.
For musical heritage, metadata allows for the preservation and dissemination of musical works and traditions, but also aid in the research and study of music history and culture \cite{giannoulakis2018metadata}.
When integrating both views, metadata can help to promote diversity and inclusivity in the music industry by highlighting lesser-known genres and artists, while integrating information and artefacts of cultural interest \cite{de2021introduction}.

Hence, a model that can consistently describe metadata is highly desirable -- as it enables linking entities and concepts from various datasets (e.g. a composer is linked to a tune that has no authors in another collection).
Semantic Web technologies can help achieve interoperability, as they facilitate data access and integration, resource discovery, semantic reasoning and knowledge extraction \cite{berners2001semantic}.
In the Resource Description Framework \cite{lassila1998resource}, data is described as \texttt{<subject-predicate-object>} triples using ontologies, and released as Knowledge Graphs (KGs).

To achieve interoperability, one possibility akin to \cite{corthaut2008connecting} is to let stakeholders design their own domain-specific ontologies, then use alignment algorithms to find connections between them (e.g. \texttt{MusicalWork} and \texttt{Composition} referring to the same concept).
However, this approach comes with three major drawbacks: (i) ontology alignment is error-prone, hence links would still require manual inspection; (ii) even when alignment is sound, the semantics of classes and relationships may vastly differ across domains, which in turn, may create inconsistent alignments; (iii) it does not address the problem in the long-term.

\subsection{Challenges and requirements for interoperability}\label{ssec:intro-challenges}

Another possibility is to reuse current ontologies for music metadata, such as the Music Ontology (MO) \cite{raimond2007music} and the DOREMUS ontology \cite{achichi2015doremus}.
However, modelling music metadata across different genres and historical periods, to accommodate various use cases over heterogeneous data sources poses a number of challenges.
First of all, it requires a perspective that harmonises all requirements from different stakeholders -- to design a model that can be tailored to different data sources rather than to a single type of dataset.
We categorise the main challenges and requirements for metadata interoperability as follows.

\subsubsection{Domain specificity hampers interoperability}\label{sssec:challenge-domain}
% Point A: specificity hampers reuse : we need to zoom out and give possibility to zoom in to create an extendible lingua franca
When looking at current ontologies, MO leans towards modelling discographic data with a focus on contemporary music, whereas DOREMUS is inherently rooted in classical music.
These ontologies have been demonstrated to model metadata from MusicBrainz and BBC Music \cite{raimond2012evaluation}, and from classical music libraries and radio broadcasts for concerts programming \cite{choffe2016doremus}, respectively.
Their specificity makes them appealing when downstream applications show considerable overlap in terms of requirements and data.
Examples include the reuse of MO in the WASABI project \cite{buffa2021wasabi}, to support the semantic annotation of audio music (emotions, lyrics, structures), but also for music recommendation \cite{rodriguez2015ontology} and listening \cite{adamou2019crowdsourcing}; and the adoption of DOREMUS by \textit{Philarmonie de Paris}, \textit{Biblioteque National de France}, and \textit{Radio France}.

Nevertheless, when drifting from discographic data and classical music, or attempting to reuse both models, addressing e.g. cultural heritage requirements while fostering interoperability becomes difficult.
% , but also originates inconsistencies when ... those models  describing music data that suit  ... but raises concerns for interoperability
Indeed, a model reflecting the view and the interpretations ascribable to a musical genre, stakeholder, or dataset type may be difficult to reuse and extend to other domains.
For instance, a music artefact may originate from oral transmission or be the result of a creative process that does not necessarily entail a formal composition process.
The latter is common in songwriting, but also in folk music whenever a set of tunes (collected from different manuscripts) allows for the identification of a tune family \cite{van2016meertens}.
Similarly, when expressing relationships between musical artefacts (alias derivations), it is important not to impose any modelling bias that may constrain possible interpretations (e.g. an arrangement having proper musical identity vs simply providing a different instrumentation).
This is commonly referred to as ``dominance of concept'' \cite{choffe2016doremus}, whose definition should be left to users depending on their data and domain expertise.
% Rather than permeating into the the model

% Other points to consider here: relationship between composition and score, composition emerging as a tune family, rules based on the dominance of concept (if something is the same work or not)
Rather than attempting to achieve consensus on musical concepts and jargon, accounting for the interoperability calls for an abstraction layer for music metadata (``\textit{zoom-out}'') that can then be specialised, extended, and adapted to address domain-specific requirements (``\textit{zoom-in}'').

\subsubsection{Expressivity is needed at different levels}\label{sssec:challenge-modularity}
% Point B: Expressivity at different degrees (depending on the data) ; mention that the model is hierarchical? Provide an example
Another requirement for interoperability and reuse across various data sources is providing expressivity at different degrees, i.e. the possibility to conveniently describe music metadata at the right level of detail.
For example, one data source may have granular/detailed information that requires high semantic expressivity (a composition process spread over different time, places, and involving more artists); whereas others may have basic (only the name of an artist is known) or even incomplete and uncertain information (a composition tentatively attributed to an artist).

Here, the WikiProject Music\footnote{\url{https://en.wikipedia.org/wiki/Wikipedia:WikiProject_Music}} has been successful in providing expressivity to represent music metadata from different sources.
As an extreme case of ontological flexibility, the schema underlying Wikidata -- an open-ended, multi-domain KG built collaboratively like Wikipedia -- is not specified in a previously agreed ontology, and the high expressivity overly adds complexity to the model.
This is due to Wikidata's scope being the most general.

% Therefore, a tradeoff between expressivity and complexity exists -- with a highly expressive model potentially jeopardising its understanding and reuse [CT].

\subsubsection{Provenance is fundamental for data integration}\label{sssec:challenge-provenance}
% Point B Provenance of metadata can be incomplete, ambiguous
Accounting for provenance is a central requirement for both cultural heritage and music industry.
This becomes fundamental when integrating Knowledge Graphs from different datasets and stakeholders -- as every single bit of data (each triple) should be attributable to a dataset/KG.
Furthermore, integrating provenance is also needed within the context of a single dataset, at least for claims and links.

\textbf{Claims-Interpretations}. Cultural heritage applications often require representing debatable statements or claims \cite{daquino2020knowledge, daquino2022expressing}.
These are usually the result of an interpretation process based on factual or documentary evidence (a dataset, a manuscript, etc.), and following a methodology and/or theory.
Examples include personal information (e.g. the year/place of birth of a composer), and authorship claims (e.g. a composition being attributed to an artist).

\textbf{Links and identifiers}.
% Representing links and identifiers to external resources and datasets is another common requirement in the music domain.
These includes links to artists' official websites, fan pages, discussion forums, music reviews, record shops; as well as identifiers from music databases (e.g. MusicBrainz, Discogs, AllMusic), streaming platforms (e.g. Deezer, Spotify), and authoritative sources (e.g. ISNI, ISWC, ISRC).
As most links and identifiers are crowdsourced or automatically inferred by entity linking algorithms, modelling provenance here promotes traceability and accountability of data sources.

Notably, Wikidata addresses both these requirements, as every triple is considered a statement per se, for which so-called \textit{references} can be appended and ranked.
References may include information on the source, whether a computational method was used, and a date of retrieval.

% Point C: FRBR is difficult and heavy : EDIT to related work
% Difficulties of FRBR models (from the other article rejected)

% \todo[inline]{But this is also a challenge in SW (text from Aldo)}
% In addition, non-textual information entities such as music, images and videos have a limited representation counterpart in the Semantic Web: they are usually referred at the level of information objects, but actual content (realisations of music, pictures, videos, etc.) is mostly referenced as traditional Web 1.0 URIs, and not frequently annotated with semantic URIs denoting information objects.
% This complex situation reveals a substantial limitation in the practical integration of semantic data and content on the open Semantic Web, which leaves to proprietary integration (iTunes and YouTube to mention the largest ones) the main operational implementation of multimedia semantics.

\subsection{Our contribution}

We leverage the expertise and complementary views of various music stakeholders (musicologists, data engineers, music analysts, and heritage archivists) to contribute:

\begin{itemize}
    \item The \emph{Music Meta} ontology, a rich flexible model to describe Western music metadata and its provenance at different levels of granularity. % -- depending on the type and specificity of the data at hand.
    \item An example-driven validation of the model, focused on the data elicited from four different stakeholders. 
    \item Code support to create Music Meta KGs without expert knowledge of the model, with automatic alignments to the MO, DOREMUS, and Wikidata.
\end{itemize}

% Our model is intuitive, modular, and designed using state of the art methodologies for ontology engineering ( based on ontology pattern.

% Related to the former point, our queries are not complex, compared to FRBR extensions

  % - Similarly to DOREMUS, we address metadata requirements across different stakeholders... and contribute towards a unified/universal model of music metadata (aware that more work is needed to achieve this vision)
  
  % - We have full alignment to the MO and DOREMUS, which means that new users can ... while ... interlinking existing datasets and KGs in this domain (CALMA, WASABI, LED, etc.)

\section{Related work}\label{sec:related}
% Some music ontologies on music content / contexts
Besides metadata, the use of Semantic Web technologies in the music domain has contributed several ontologies, covering a variety of musical aspects and spanning both symbolic and audio music.

% 1. MUSIC THEORY and NOTATION
Among them, the Music Theory Ontology \cite{rashid2018} describe theoretical concepts of compositions, whereas the Music Score \cite{jones2017} and Music Notation \cite{cherfi2017} propose granular ontologies to represent elements of music scores.
OMAC expresses features of musical entities but also musicological claims \cite{sanfilippo2022musicclaims}, while the OMRAS project \cite{fazekas2010} contributed ontologies to describe music chords as well as concepts related to tonality and temperament.

% 2. Implicitly or explicitly AUDIO-BASED
In the audio domain, ontologies describe music production \cite{fazekas2011}, audio features \cite{allik2016}, effects \cite{wilmering2013}; an also model listening habits/taste \cite{celma2008foafing}, music-induced emotions \cite{song2009emotions}, music structure \cite{fields2011, harley2015}, and musical similarities \cite{jacobson2009similarity}.
%Higher-level features related to music structure are also modelled in the Segment \cite{fields2011} and CHARM \cite{harley2015} ontologies.

% 3. From specific to inter-operable efforts
These ontologies have specific focus, and many were developed as stand-alone projects, with little or no alignment \cite{carriero2021semanticintegration}.
Instead, some ontologies focus on achieving interoperability between notations, taxonomies, and formats.
These include the Internet of Musical Things \cite{turchet2020internet}, where heterogeneous musical objects are envisioned to coexist; the Music Annotation Pattern \cite{poli2022muanp} which allows to model music annotations in the JAMS format \cite{humphrey2014jams}; and the Hamse ontology \cite{poltronieri2022hamse} describing musical features for musicological research.
Similarly, \cite{lewis2022arrangements} models abstract annotations of musical works, rather than concrete encodings.

Interoperability at the level of musical content level resulted in successful MIR applications, such as the MIDI Linked Data Cloud \cite{merono2017} -- integrating MIDI music to learn embeddings over the resulting KG \cite{lisena2022midi2vec}; and ChoCo \cite{polifonia2023choco} -- a chord corpus integrating 18 chord datasets and enabling novel workflows for computational creativity \cite{polifonia2023harmory}.

\section{The Music Meta Ontology}\label{sec:mmeta}

\begin{figure*}[t]
    \centering
    \includegraphics[width=\linewidth]{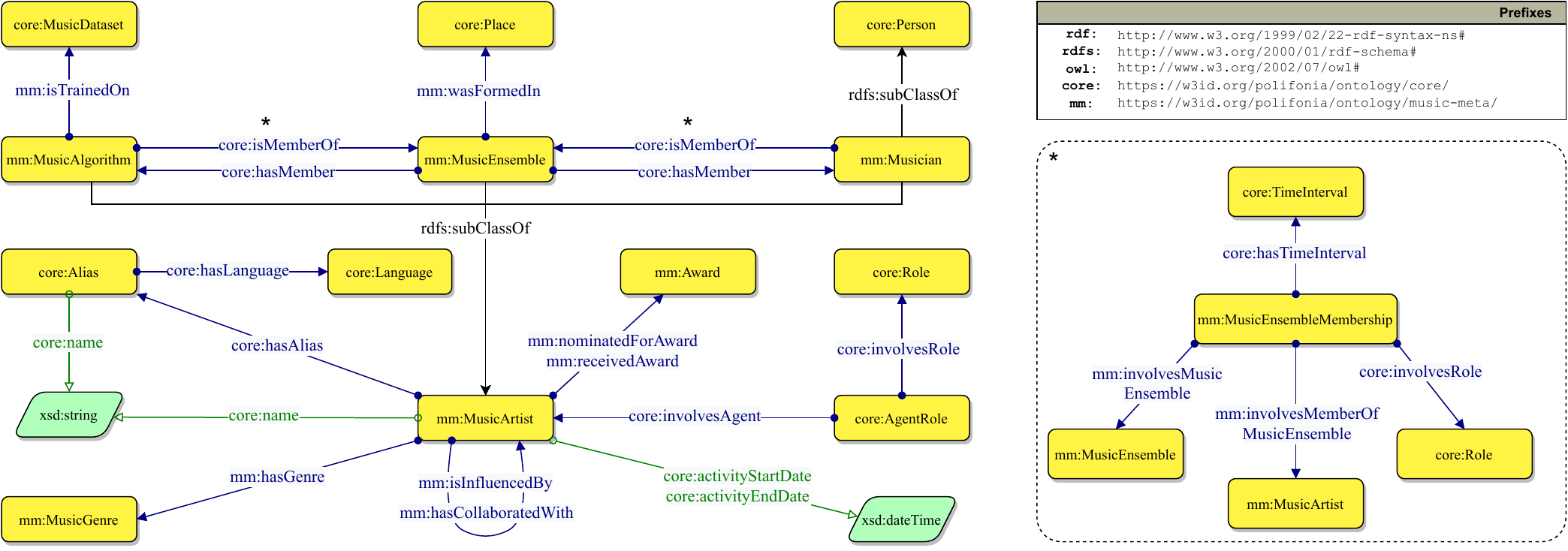}
    \caption{Describing music artists as musicians, music ensembles, and algorithms using the Graffoo notation (\textit{yellow} boxes are classes, \textit{blue/green} arrows are object/datatype properties, \textit{purple} circles are individuals, \textit{green} polygons are datatypes).}
    \label{fig:music-agents}
\end{figure*}

To derive requirements from various music stakeholders, we leverage the domain expertise and views in Polifonia -- a European H2020 project aiming to connect ``music, people, places and events'' from the 16th century.
The interdisciplinarity of Polifonia, involving data engineers, anthropologists, ethnomusicologists, historians of music, linguists, musical heritage archivists, cataloguers, and creative professionals -- makes it an ideal testbed for this work.

Music Meta is part of the Polifonia Ontology Network \cite{polifonia2023pon}, from which we reuse the CORE module.
This is done to consistently reuse general-purpose elements of design (e.g. Person, Time, Place) and ontology design patterns.
The reuse of this module also ensures alignment with other foundational models (FOAF, Dublin Core, etc.).

The ontology (prefixed as \texttt{mm}) is available at the following URI: \texttt{\url{https://w3id.org/polifonia/ontology/music-meta/}}, and is released as open source project under the \texttt{CC-BY 4.0} on GitHub\footnote{\url{https://github.com/polifonia-project/music-meta-ontology}}.
%which includes documentation, data examples, and a library for the data transformation \footnote{\url{Link-omitted-for-double-blind-review.}} (c.f. Section~\ref{ssec:mmeta-library}).

\subsection{Methodology}\label{ssec:methodology}

% Succinct overview of the XD Methodology (stories, CQ)
The development of Music Meta is driven by eXtreme Design (XD) \cite{blomqvist2016engineering}, an agile ontology engineering methodology that makes extensive use of ontology design patterns (ODPs) -- small ontologies that work as reusable templates for recurrent modelling problems.
An ODP is intuitive and compact, clearly and formally defined, tackles a specific (sub)set of requirements, and is designed for a modular reuse, enabling a pragmatic cognitive analysis \cite{ReuseLandscape2020}.

In XD, a story-based approach guides the collection of requirements.
A story is a framework for \emph{customers} to describe their needs, and is composed of 4 sections: (i) the persona, a description of a typical user; (ii) the overarching goal they need to address; (iii) the scenario, describing how the goal will be address; (iv) the competency questions (CQs) translating needs into formal requirements.
Ontology modelling starts iteratively from the CQs, and is based on the reuse of ODPs and existent templates.
% entities extracted from existing ontologies as templates.

\subsubsection{From FRBR to Information Objects/Realisations}
At the core of Music Meta lies the use of the Information-Realisation (IR) ODP \cite{infoRealizationODP}.
An \emph{information object} is a non-physical social object carrying information that can have one or multiple materialisations (\emph{information realisations}).
Each realisation is a particular physical object, or event, realising the the \textit{information object}, or involving the latter as a participant.
Both information object and realisation are intended as information entities (IE), i.e. (social) objects created and/or used
to communicate, reason, and specify new entities.
This allows to distinguish between a piece of information (e.g. the \emph{content} of a composition) from how it is materialised (e.g. as a performance).

On the other hand, both the Music Ontology \cite{raimond2007music} and DOREMUS \cite{choffe2016doremus} are built on top of different flavours of FRBR\cite{frbr} (FRBRer and FRBRoo, respectively).
FRBR is a conceptual model describing bibliographic resources at four levels: \textit{Work}, \textit{Expression}, \textit{Manifestation}, and\textit{Item}.
In contrast, the two levels of the IR pattern map to \textit{Expression} and \textit{Item}, since \textit{Work} and \textit{Manifestation} are said to provide non-informative conceptualisations \cite{infoRealizationODP}.
Moreover, \cite{riley2008frbr} argues that FRBR's Works -- intended as ``entities that pre-exist expressions'', cannot represent improvisations or traditional music, as they do not derive from a formal composition process leading to a realisation.
FRBR's Work is often ambiguously intended as an entity retrospectively created for grouping multiple expressions for cataloguing needs.
As for the Manifestation level, while its representation is straightforward in the bibliographic domain (e.g. the printed version of a book), its correspondence in the music domain is not fully intuitive, as it may relate to either a recording, a score, a compact disc, or all the above -- thereby introducing complexity and ambiguity.

Nevertheless, being aligned to two levels of FRBR, the IR ODP makes our model leaner and flexible, while still achieving interoperability with FRBR-based (music) ontologies.
In fact, IE patterns are meant to boost the semantic integration of contents, tools, platforms, resources that are silo-ed or non-interoperable \cite{infoRealizationODP}.

% Say that they both rely on FRBR...
% Whereas MO is based on a ``mild'' reuse of FRBR for the central components in the model (MusicWork, MusicExpression, MusicManifestion, and MusicItem), DOREMUS heavily rely on FRBRoo to the largest degree.
% Although the latter has large expressivity, the strong connection with FRBR adds complexity to the model in different ways: (i) query become complicated; (ii) the model is hard to explain and document.

\subsection{Main elements of design}

From Polifonia's CQs\footnote{\url{https://github.com/polifonia-project/stories}}, we identified those related to metadata, and aimed for a model capable to address the requirements in Section~\ref{ssec:intro-challenges}. 
Music Meta follows a hierarchical design (where each level extends the former to add expressiveness) and is complemented by data transformation rules to conveniently translate one level into another.
%\todo[inline]{Move Andrea's selection of CQ here as a table.}

To enable data integration from existing knowledge bases and datasets, we align Music Meta to other ontologies: the Music Ontology, DOREMUS, and Wikidata, after having identified common/similar classes and properties.

\subsubsection{Music artists}

To represent music creatives the class \texttt{mm:MusicArtist} generalises over musicians (\texttt{mm:Musician}), ensembles (\texttt{mm:MusicEnsemble}), and computational methods (\texttt{mm:MusicAlgorithm}), as illustrated in Figure~\ref{fig:music-agents}.
Musicians are seen as a specialisation of persons who can optionally be associated to a medium of performance (e.g. voice, guitar), and be part of a music ensemble (e.g. \texttt{MusicGroup}, \texttt{Orchestra}, \texttt{Choir}).
Depending on the data available, the latter can be expressed either through a membership relationship (\texttt{core:isMemberOf}), a specialisation of the former, such as \texttt{mm:isSingerOf}, or through a \texttt{mm:MusicEnsembleMembership} when the period of participation of the musician is available.

All music artists can be associated to (one or more) \texttt{mm:MusicGenre}(s), express influences or collaborations, and share a period of activity.
Here, the start date refers to the foundation for music ensembles, whereas the end date is used for discontinued projects for algorithms.

\subsubsection{Music inception}\label{sssec:mmeta-inception}

% \todo[inline]{The most critical one -- as we should should show 1.1.1 (abstraction) and 1.1.2 (hierarchical design). Starting with the full diagram as we have here.}
\begin{figure*}[h]
    \centering
    \includegraphics[width=\linewidth]{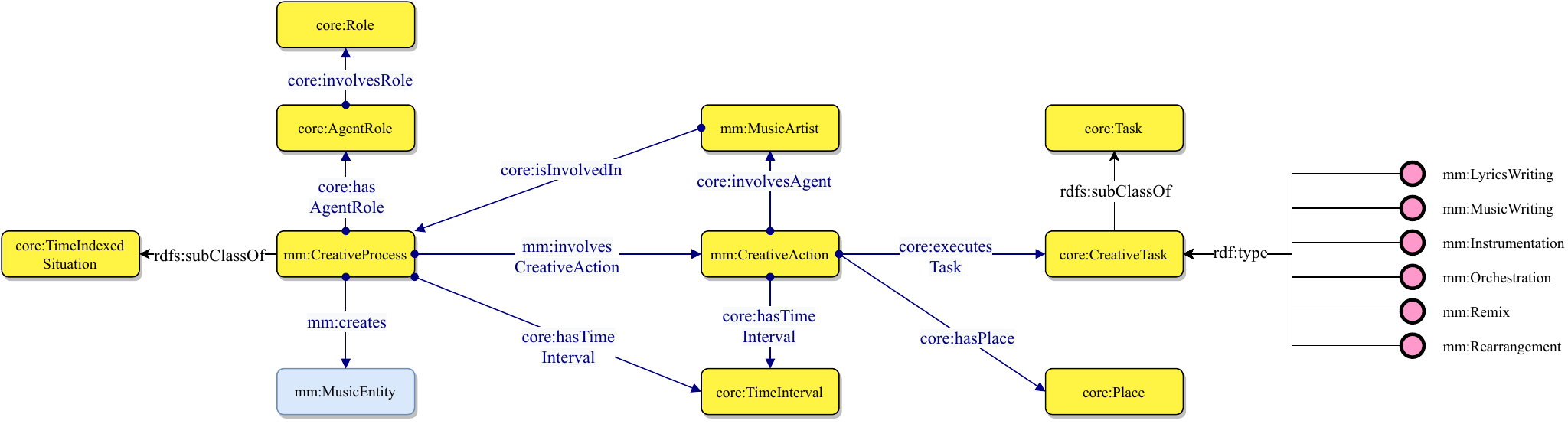}
    \caption{Abstracting music inception as an product of a creative process, involving music artists in activities (music writing, instrumentation, etc.), defined in time and space and according to different roles.}
    \label{fig:music-creation}
\end{figure*}

\begin{figure*}[htbp]
    \centering
    \includegraphics[width=\linewidth]{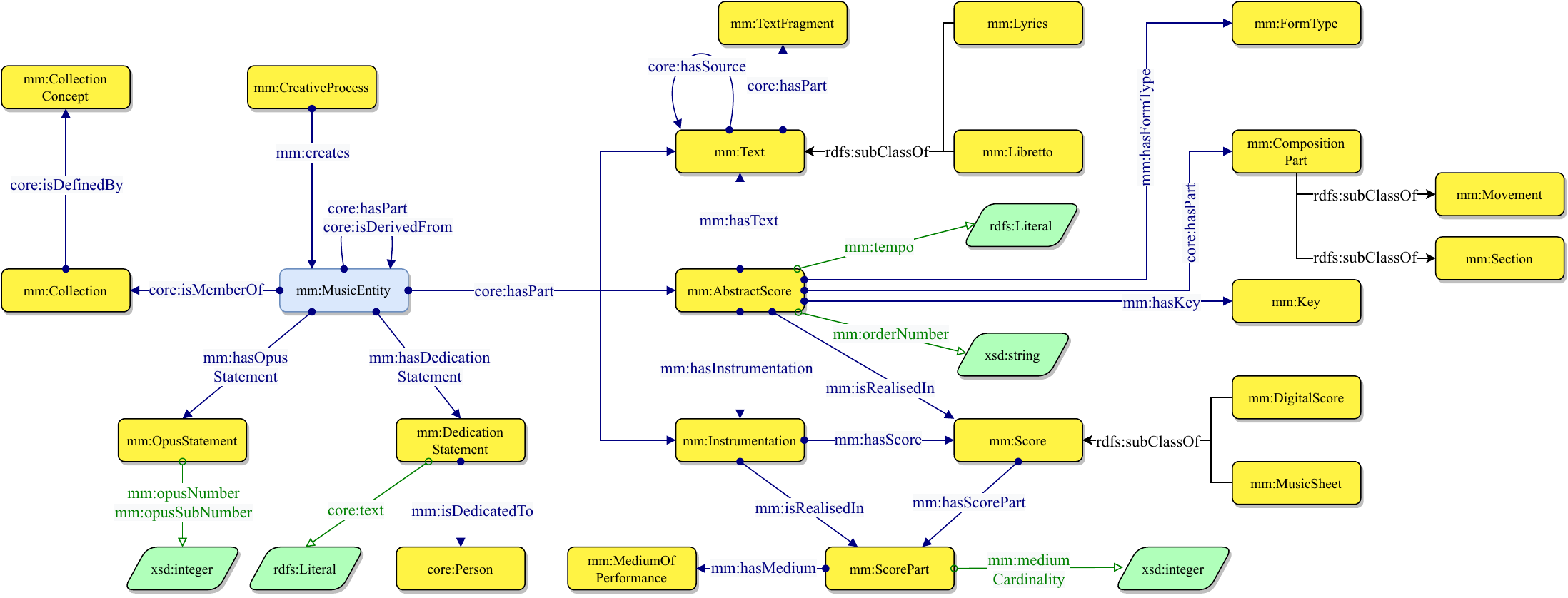}
    \caption{Describing a music entity and the elements it contains: Text, AbstractScore and Instrumentation.}
    \label{fig:music-entity}
\end{figure*}

The focal point of Music Meta is the \texttt{mm:MusicEntity} class (Figures \ref{fig:music-creation} and \ref{fig:music-entity}). 
This class represents an Information Object, which is defined as the sum of all the elements that make up a piece of music.
A Music Entity is composed of several components, including lyrics (generalised through \texttt{mm:Text} to also account for \texttt{mm:Libretto}), the entailed musical content (\texttt{mm:AbstractScore}) and its instrumentation (\texttt{mm:Instrumentation}).

% An abstract concept that allows to refer to the actual musical content of a MusicEntity. This makes it possible to describe musical properties that imply the existence of a hypothetical score, which may not be necessarily, or formally, materialised by the creators of the musical entity.
A \texttt{mm:AbstractScore} provides an abstraction to describe the musical properties of an entity, such as the form of a piece (\texttt{mm:FormType}), its constituents parts (e.g. \texttt{mm:Movement} or \texttt{mm:Section}), and its key (\texttt{mm:Key}).
Datatype properties also describe the tempo of the composition (\texttt{mm:tempo}) and its order (\texttt{mm:orderNumber}).
A \texttt{mm:Instrumentation} can instead be formalised in a \texttt{mm:Score}, which can be either digital or paper.
Through the score, the instrumentation describes one or more \texttt{mm:MediumOfPerformance}, each of which has a cardinality (e.g. 3 violins).

It is also possible to describe relationships between different Music Entities, defined by parthood (\texttt{mm:hasPart}) and derivation (\texttt{mm:isDerivedFrom}).
Derivations are used at the user's discretion, based on the dominance of concept \cite{choffe2016doremus} (whose criteria attribute proper identity to a musical entity) and can be of different types: revision, transposition, cover, reconstruction, reduction, etc.
This makes it possible to describe different types of compositions, rearrangements and modifications of an original piece, as well as influences and more complex types of derivations.
For example, the production of a cover song (e.g. in a different musical genre) may keep the lyrics and introduce a new composition and instrumentation, hence resulting in a new \texttt{mm:MusicEntity}.
In addition, Music Entities can be organised in \texttt{mm:Collection}, according to a \texttt{mm:CollectionConcept} that binds them together.

In sum, the model provides flexibility across periods and genres as the proposed classes allow generalisations to be made about the text, the musical composition and its arrangement. (c.f. Section~\ref{sssec:challenge-domain}).
Through the specialisation of classes, depending on the target domain/application, specificity can easily be achieved (c.f. Section \ref{sssec:challenge-modularity}).
For example, a tune family can be seen as a \texttt{mm:Collection} encompassing several tunes (as music entities) based on specific criteria (e.g. similarity, provenance).

\subsubsection{From performance to recording and broadcast}\label{sssec:mmeta-audio}

\begin{figure*}[t]
    \centering
    \includegraphics[width=\linewidth]{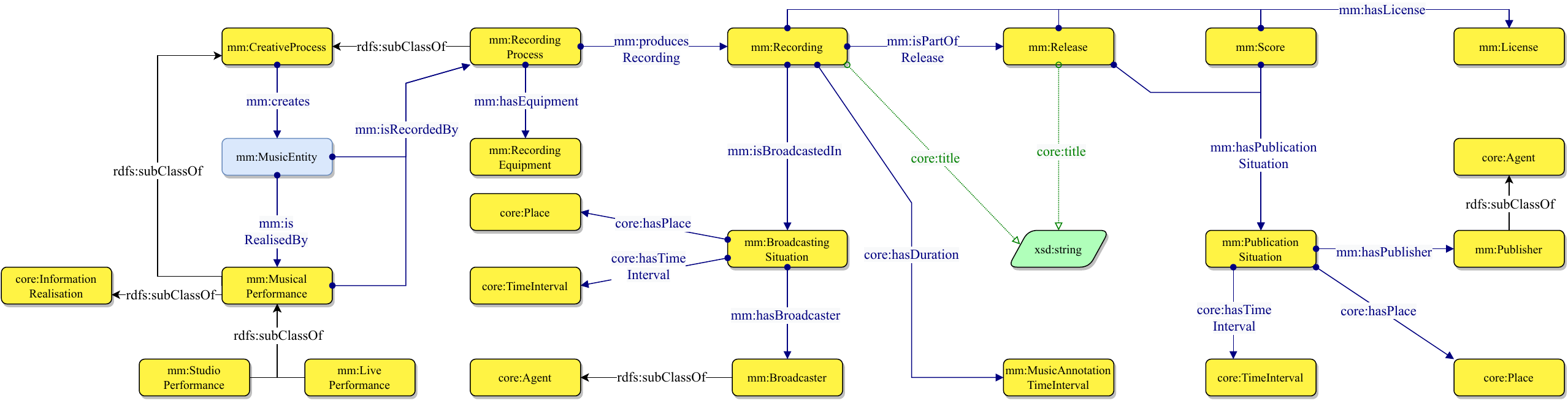}
    \caption{Describing performance, recording, broadcasting, publication, and licensing.}
    \label{fig:music-performance}
\end{figure*}

The realisation of a \texttt{mm:MusicEntity} is exemplified by \texttt{mm:MusicalPerformance}, which can be either live (\texttt{mm:LivePerformance}) or in a studio (\texttt{mm:StudioPerformance}).
As illustrated in Figure~\ref{fig:music-performance}, the place and time interval of a performance are described by \texttt{core:Place} and \texttt{core:TimeInterval} -- involving one or more music artists (optionally, with a specific role).
A performance may also create a new \texttt{mm:MusicEntity} if, e.g., the execution differs significantly from the original version.

A Music Entity can also be recorded by means of a \texttt{mm:RecordingProcess}, which is a subclass of a \texttt{mm:CreativeProcess}.
%that allows for specifying location, time interval and persons involved in recording the song. 
This makes it possible to describe information about both the production (e.g., producers) and the technical aspects of it (e.g., sound engineer, equipment used).
The recording process produces a \texttt{mm:Recording}, which is contained in a \texttt{mm:Release}.

Information about the broadcasting of a recording is modelled through the \texttt{mm:BroadcastingSituation} class (an instance of the Situation ODP \cite{gangemi2002dolce}), which describes when and where the song was broadcast, and by which broadcaster (\texttt{mm:Broadcaster}).

\subsubsection{Publishing and licensing information}\label{sssec:mmeta-pub}
% \todo[inline]{This is for music entities, recordings, releases, etc. It is important to support License as MO (I think we forgot? But it is in our meta-CQs.)}

The \texttt{mm:PublicationSituation} class describes information about the publication of a release, which is common to the publication of a \texttt{mm:Score} (c.f. Figure~\ref{fig:music-performance}). 
For both a release and a score, it describes when and where they were published, and by a \texttt{mm:Publisher}.

Licence information is described by the \texttt{mm:License} class, which applies to records, releases and scores.

\subsubsection{Modelling links and integrating provenance}\label{sssec:mmeta-prov}

\begin{figure}[h]
    \centering
    \includegraphics[width=\linewidth]{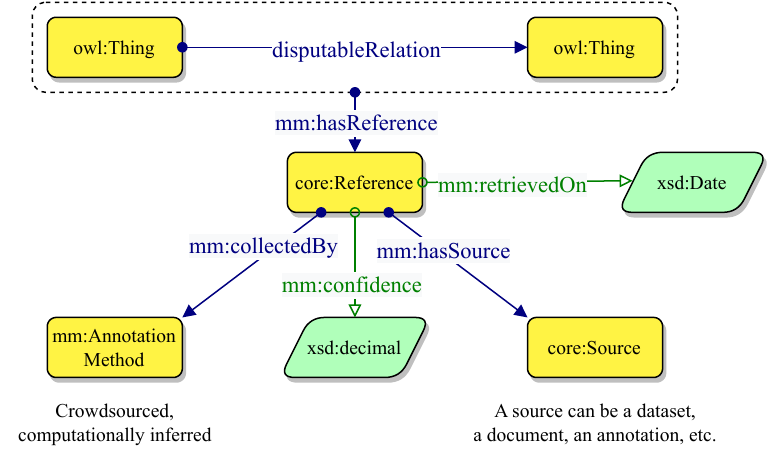}
    \caption{Our pattern to describe provenance with RDF*.}
    \label{fig:rdfstar-prov}
\end{figure}

We propose a pattern based on \emph{RDF*} \cite{hartig2017foundationsstar} to describe the provenance at different levels (Figure \ref{fig:rdfstar-prov}).
The use of RDF* is particularly useful for this purpose, as it allows to embed provenance information to every triple in the dataset.
This simplifies and streamlines the model, eliminating the need for n-ary relations or reification for each triple.

The proposed pattern is straightforward and comprises the class \texttt{core:Reference}, which describes the source of the reference (using the class \texttt{core:Source}) and the method used to obtain the annotation (using the class \texttt{core:SourceMethod}).
Additionally, the datatype properties \texttt{core:confidence} and \texttt{core:retrievedOn} describe the confidence of the annotation and the date it was produced, respectively.

\subsection{Conversion rules and code support}\label{ssec:mmeta-library}

To facilitate the reuse of Music Meta and its data conversion into OWL/RDF Knowledge Graphs, we developed \texttt{PyMusicMeta} -- a library to map arbitrary music metadata into RDF triples.
This enables a practical and scalable workflows for data lifting to create Music KGs without expert knowledge of our ontological model.
The library is developed in Python as an extension of RDF-Lib \cite{bottinger2018rdflib}.

% \texttt{PyMusicMeta} allows for the creation of RDF triples from textual data, offering the advantage of easy data generation using our model.
% \texttt{PyMusicMeta} provides different methods for adding triples to a graph, ...
With each triple, \texttt{PyMusicMeta} adds alignments to the supported schema whenever possible.
For example, the pseudo triple \texttt{<DavidBowieURI, rdf:type, mm:Musician>} in Music Meta will be complemented with \texttt{<DavidBowieURI, rdf:type, http://purl.org/ontology/mo/MusicArtist>} for Music Ontology, \texttt{<DavidBowieURI, rdf:type, http://erlangen-crm.org/E21\_Person>} for DOREMUS (via the Erlangen Conceptual Reference Model \cite{merges2012erlangen}) and \texttt{<DavidBowieURI, rdf:type, https://www.wikidata.org/wiki/Q639669>} for Wikidata; to achieve interoperability of the Music KG.

% Additionally, an automated version of the library is available, allowing users to input a JSON file formatted according to the documentation, and converting it directly to a KG.
%The library will then automatically transform the data contained in the JSON into RDF/OWL.
%If the JSON file is missing any fields from the template file, the corresponding triples will simply not be generated.

\section{Validation and Adoption}\label{sec:validation}

Following the XD methodology (c.f. Section~\ref{ssec:methodology}), we validate Music Meta against the competency questions (CQs) driving its design.
In this context, testing consists in formulating logical statements for each competency question -- using the ontology as a formal model.
Logical statements are encoded as SPARQL queries to evaluate the model.
Examples of tested CQs include ``\textit{In which time interval did the creation process take place?}'' and ``\textit{Which is the language of the name/alias of a music artist?}''.
The complete list of CQs, together with their correspondent SPARQL queries can be found in the project's repository.
This also contributes a test framework where the ontology is automatically tested using the available SPARQL queries \cite{fernandez2017ontology}, whenever changes occur or new requirements are supported in future versions of Music Meta.

% The evaluated CQs are categorised based on these desiderata:
% (i) Heterogeneity: the ontology must describe pieces of music from different genres and historical periods;
% (ii) Composibility: the ontology has to describe pieces of music organised in various collections, including catalogued collections and those defined through musicological analysis, such as corpora;
% (iii) Derivation: the model must describe the various ways in which a piece of music could be derived or rearranged, as well as multiple performances of the same composition, where the performing act acts as a well-defined creative process;
% (iv) Dissemination: the ontology needs to describe the broadcasting of a music track, as well as licensing;
% (v) Provenance: every triple should be traced back to its source and the person/entity who asserted it.

% Having each competency questions successfully mapped to a SPARQL query confirms that ...
Music Meta has already been used in ChoCo \cite{polifonia2023choco}, the largest Harmony KG to date, obtained from the integration of 18 MIR datasets\footnote{\url{https://github.com/smashub/choco}}.
The ontology has also been \emph{specialised} for folk metadata (Tunes Ontology) and \emph{extended} to describe music datasets (CoMeta Ontology).
All ontologies are part of the Polifonia Ontology Network (PON) and can be found at \url{https://github.com/polifonia-project/ontology-network}.
We also provide documentation, examples, and tutorials\footnote{\url{https://polifonia-project.github.io/ontology-network/}}.

%\begin{description}
%    \item[CQ1] \textit{Has composition X been identified as variant in a tune family?}
%    \item[CQ2] \textit{What are alternative titles for composition X?}
%    \item[CQ3] \textit{What licence applies to the metadata? To the digital score? To the audio?}
%    \item[CQ4] \textit{Who published and broadcast the recording?}
%    \item[CQ5] \textit{In which collections/datasets does song X occur?}
%    \item[CQ6] \textit{Where was the music printed source published?}
%    \item[CQ7] \textit{In which collections/datasets does song X occur?}
%    \item[CQ8] \textit{For a given track, which are its performances and where did they took place?}
%\end{description}

% \cite{poli2022muanp}
\section{Conclusions}\label{sec:conclusions}

The interoperability of metadata is an essential requirement for the integration of music datasets, which is currently hampered by the specificity of existent ontologies.

Our work addresses interoperability requirements for the design of the Music Meta ontology -- a rich and flexible semantic model for (Western) music metadata across different genres and periods, for various stakeholders and music datasets.
The model is based on the Information-Realisation ontology design pattern, allowing to reduce complexity while maintaining alignment to other ontologies (Music Ontology, DOREMUS).
We validate Music Meta following the XD methodology, to demonstrate the support of requirements collected from various stakeholders (music analysts, archivists, musicologists, and data engineers).
The model has modular design -- allowing users to describe music data depending on their specificity and type, while providing provenance support through RDF*.

We are extending the evaluation of Music Meta across cultural heritage and music industry datasets, while working with our stakeholders to specialise the model for the integration and release of Music Knowledge Graphs.

\newpage
\textbf{Acknowledgements} This project has received funding from the European Union’s H2020 research and innovation programme under grant agreement No 101004746. The authors also acknowledge Philippe Rigaux, Peter van Kranenburg, Marco Gurrieri, and Mari Wigham for their support and feedback throughout the design of Music Meta.

% For bibtex users:
\bibliography{bibliography}

\end{document}